 \documentclass[preprint2]{aastex}

\def\lsim{\mathrel{\rlap{\lower 4pt \hbox{\hskip 1pt $\sim$}}\raise 1pt \hbox
        {$<$}}}
\def\gsim{\mathrel{\rlap{\lower 4pt \hbox{\hskip 1pt $\sim$}}\raise 1pt \hbox
        {$>$}}}

\slugcomment{Submitted to the Astrophysical Journal}

\shorttitle{Peculiar Chemical Abundances in the Starburst Galaxy
M82 \\and Hypernova Nucleosynthesis}
\shortauthors{Umeda \& Nomoto}

\begin{document}
\title{Peculiar Chemical Abundances in the Starburst Galaxy
M82 \\and Hypernova Nucleosynthesis}

\author{Hideyuki Umeda\altaffilmark{1}, 
Ken'ichi Nomoto\altaffilmark{1,2},  
Takeshi Go Tsuru\altaffilmark{3},  
Hironori Matsumoto\altaffilmark{3}} 


\affil{$^1$
Department of Astronomy, University of Tokyo, Hongo, Bunkyo-ku,
113-0033, Japan}
\email{umeda@astron.s.u-tokyo.ac.jp}

\affil{$^2$
Research Center for the Early Universe, 
University of Tokyo, Hongo, Bunkyo-ku,
113-0033, Japan}
\email{nomoto@astron.s.u-tokyo.ac.jp}

\affil{$^3$Department of Physics, Kyoto University,
Sakyo, Kyoto, 606-8502, Japan} 
\email{tsuru@cr.scphys.kyoto-u.ac.jp}

\email{\rm 
To appear in the Astrophysical Journal 578, 2002}

\begin{abstract}
 X-ray observations have shown that the chemical 
abundance in the starburst galaxy M82 is quite rich in
Si and S compared with
oxygen. Such an abundance pattern cannot be
explained with any combination of conventional Type I and II 
supernova yields. Also the energy to heavy element mass ratio
of the observed hot plasma is much higher than the value
resulted from normal supernovae. 
We calculate explosive nucleosynthesis in core-collapse hypernovae
and show that the abundance pattern and the large ratio
between the energy and the heavy
element mass can be explained with the hypernova nucleosynthesis. 
Such hypernova explosions are expected to occur for stars more
massive than $\gsim 20-25 M_\odot$, and likely dominating the starburst,
because the age after the starburst in M82 is estimated to be
as short as $\sim 10^6 - 10^7$ yr. We also investigate pair-instability 
supernovae ($\sim 150-300 M_\odot$) and conclude that the energy
to heavy element mass ratio in these supernovae is too small
to explain the observation.
\end{abstract}

\keywords{galaxies: individual : M82 --- stars: abundances --- supernovae: general
--- galaxies: starburst --- nucleosynthesis}

\section{Introduction}

 M82 is the most active nearby starburst galaxy.  Recently several
exciting discoveries have been made for M82.  X-ray observations have
revealed the presence of intermediate mass black holes with masses 
$10^3-10^6$ M$_{\odot}$ in M82
(Matsumoto et al. 2001; Kaaret et al. 2001), whose locations were found
to coincide with the star clusters by SUBARU observations
(Harashima et al. 2002).  Also the very energetic
expanding molecular super-bubble has been discovered (Matsushita et
al. 2000).  These findings have created lots of interest in the
formation of black holes of various masses and the evolution
of star burst galaxies (e.g., Ebisuzaki et al. 2001).

 The critically important information for understanding the
evolution of star burst galaxies is the chemical abundances.  The
abundance information has also been provided by X-ray observations.
Tsuru et al. (1997) observed M82 with ASCA in the $0.5-10$ keV X-ray 
band and found that its spectrum can be fit with a three component
model: a point-like hard component and extended soft-medium components. 
 From the observed emission lines they also obtained abundances
of O, Ne, Mg, Si, S and Fe, and found that the abundance
pattern is peculiar: Si and S are much abundant than O and Fe 
compared with the solar ratio.
Also the O/Fe ratio is almost solar. Tsuru et al. (1997) 
concluded that this abundance pattern cannot be
reproduced with any combination of the previous Type Iabc and Type II
supernova (SN Iabc and SN II)
yields (Nomoto, Thielemann \& Yokoi 1984; Woosley \& Weaver 1995;
Thielemann, Nomoto \& Hashimoto 1996; Nomoto et al. 1997a,b). 
They also discussed that the energy to heavy element mass ratio of 
the observed hot plasma is too large to be of the supernova origin.

We should note that Tsuru et al. (1997) assumed
that all stars above 8$M_\odot$ have already exploded. 
However, the age of the starburst
is estimated to be $\sim 10^6$yr from the radio observation
(Matsushita et al. 2000) and $\sim 10^7$yr from the size 
of the X-ray halo observed with ROSAT 
(Strickland, Ponman \& Stevens 1997).
Only the very massive stars have exploded in such a short time. 

 Moreover, the previous 
SN II and SN Ibc yields adopted in Tsuru et al. (1997)
were obtained only for the explosion energy of $10^{51}$ erg.
Recently some massive supernovae have been found to explode 
much more energetically as ``hypernovae'' than normal SNe II
(e.g., Iwamoto et al. 1998; Nomoto et al. 2001).  Therefore,
we need to reconsider the previous conclusions by Tsuru
et al. (1997).

 In this paper we re-examine whether the abundance and energetics of M82 are
consistent with the supernova models taking account of the starburst age and 
hypernova explosions.  Because of the short age, the abundance 
is likely to be determined
mainly by relatively massive supernovae. The contributions of
high energy explosions would affect 
the energy to the heavy element mass ratio and the abundance pattern
in the galactic winds.
Recently we have found that in nucleosynthesis of hypernovae,
the Si and S abundances are much enhanced relative to O; and
also large Fe/O ratio ([Fe/O] $\sim 0$) can be realized
(Umeda, Nomoto \& Nakamura 2000; Nakamura et al. 2001).
These patterns are consistent with those observed in M82,
which has motivated us to
calculate detailed nucleosynthesis in massive
energetic core-collapse SNe for comparison with the M82 data.
We also investigate pair instability supernovae (PISNe)
of $150 - 300M_\odot$ stars, because PISNe also yields relatively 
abundant Si and S compared with O, and large amount of Fe.

\section{Abundance Pattern}

 The heavy element abundances in M82 observed with ASCA (Tsuru et al. 1997)
are shown in Figure 1 by open circles with error bars. For Ar, only
the upper limit is given. Here, the heavy element abundances are defined
by the mass fractions of elements with respect to light elements
(mostly hydrogen and helium). The data are normalized to the solar
ratios (Anders \& Grevesse 1989). We note two features of
the observational data. One is the overabundances of Si and S relative to O 
with respect to the solar ratio, i.e., [Si,S/O] $\sim 0.9$.
Here, [X/Y] $\equiv$ log(X/Y)$-$ log (X/Y) $_\odot$.
Second is such low abundances of O and Fe as [O/H]$\sim$ [Fe/H]
$\sim -$1.2.  

 Since these abundances were derived by fitting the ASCA
spectra with the two-temperature plasma model
for the hot gas in M82, one might argue
that these features are an artifact due to the too simple model
assumption. Therefore, we fitted the ASCA spectrum with a
multi-temperature plasma model in which an emission measure
of each temperature component follows a power-law in
temperature (i.e., the emission measure from the temperature $T$ is
proportional to $(T/T_{\rm max})^{\alpha}$, where $T_{\rm
max}$ and $\alpha$ are free parameters). We obtained
$\alpha=0.20\pm0.36$ and $T_{\rm max}=0.93\pm0.14$~keV as
the best-fit results, and the best-fit abundances are quite
similar to those of the two-temperature model
fitting. We thus conclude that the above two features are
not the artifact.

 To compare with the observed abundance pattern of M82,
we calculate nucleosynthesis in the SN II explosions of 20$M_\odot$
stars for various energies and metallicities. (More details on our massive
star evolution, explosion simulation and nucleosynthesis calculations
are given in Umeda et al. 2000 and Umeda \& Nomoto 2002).
Since [Fe/H] $\sim -1$, it is better to use low metal
star models to compare with the M82 abundances.

\subsection{Normal Type II Supernovae}

 First, let us compare the abundance pattern of M82 with the normal
SN II models for which the explosion energy is assumed 
to be $E_{51}\equiv E/10^{51}$erg = 1. This is the typical
energy for SNe II as estimated from SNe 1987A, 1993J, and 
1994I (e.g., Shigeyama \& Nomoto 1990; Blinnikov et al. 2000;
Nomoto et al. 1993, 1994). The ejected mass of 
Fe (mostly $^{56}$Fe produced by the $^{56}$Ni decay, and some
$^{54}$Fe and $^{57}$Fe)
depends on the uncertain ``mass-cut'', but here
we assume that the ejected $^{56}$Ni mass is 0.07$M_\odot$
as estimated for the above SNe. The mass-cut depends on
the progenitor mass and the explosion energy. 

 In Table 1, we summarize the ejected masses of selected elements
for a various set of metallicity $Z$ and the mass $M$
of SN II models with $E_{51}=1$. The abundances in the ejecta
depend on the convective efficiency $f_k$ and the 
$^{12}$C($\alpha,\gamma)^{16}$O rate adopted for the stellar
evolution models as shown in Table 1. Here, the parameter $f_k$ 
is proportional to the diffusion coefficient
of convective mixing (Spruit 1992; Umeda et al. 1999). 
In Figure 1, we plot the abundances
of these 20$M_\odot$ SNe II models for various metallicities. 
The left and right panels show the cases with relatively slow and fast
convective mixing ($f_k$=0.05 and 0.15), respectively. 
The overall abundances are normalized to the observed Si abundance.
In this section, we can compare only the theoretical and observed 
abundance ratios. The overall
normalization is constrained by the energy to heavy element mass ratios
discussed in the section 3.

\begin{figure}
\plotone{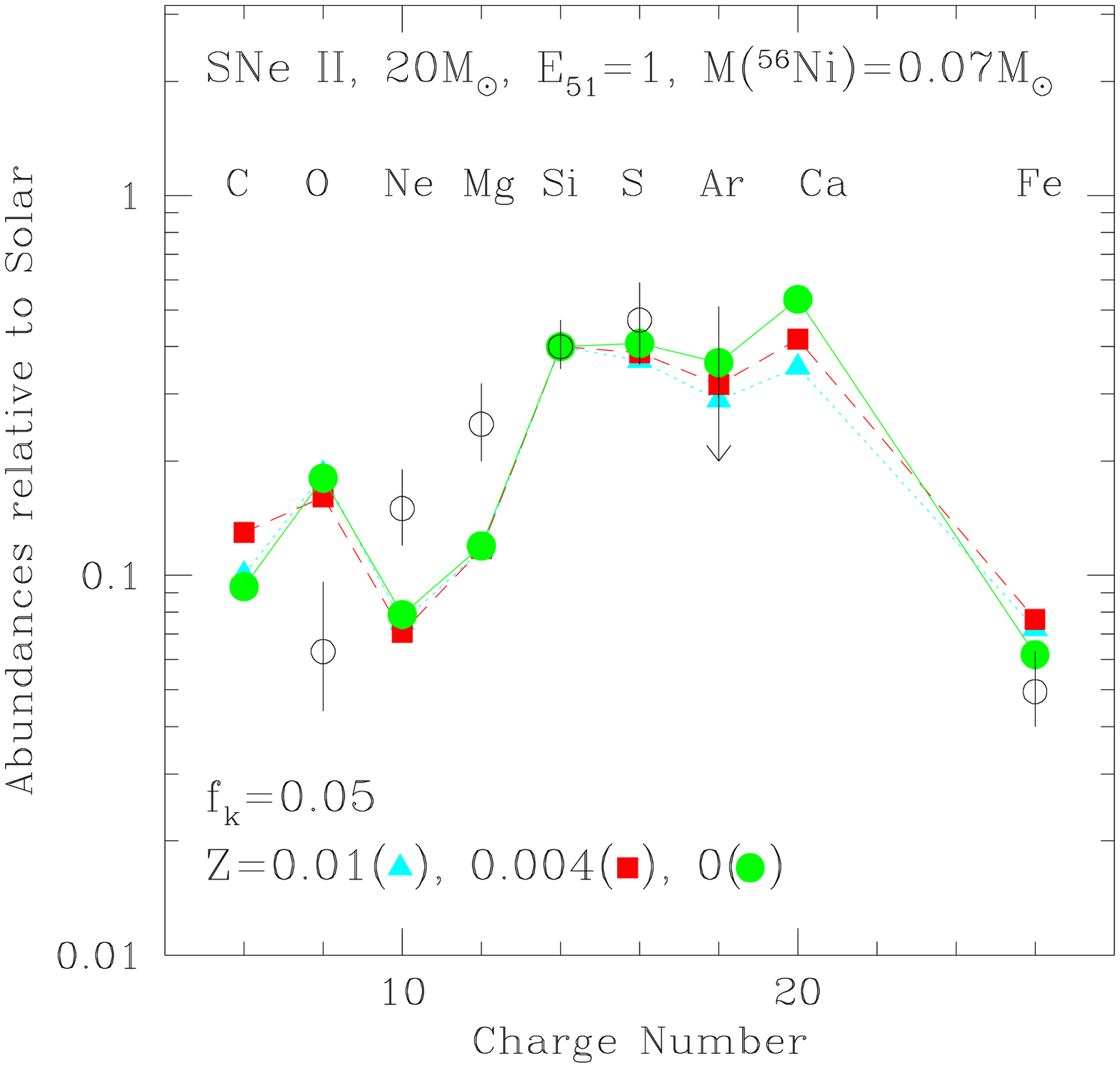}
\plotone{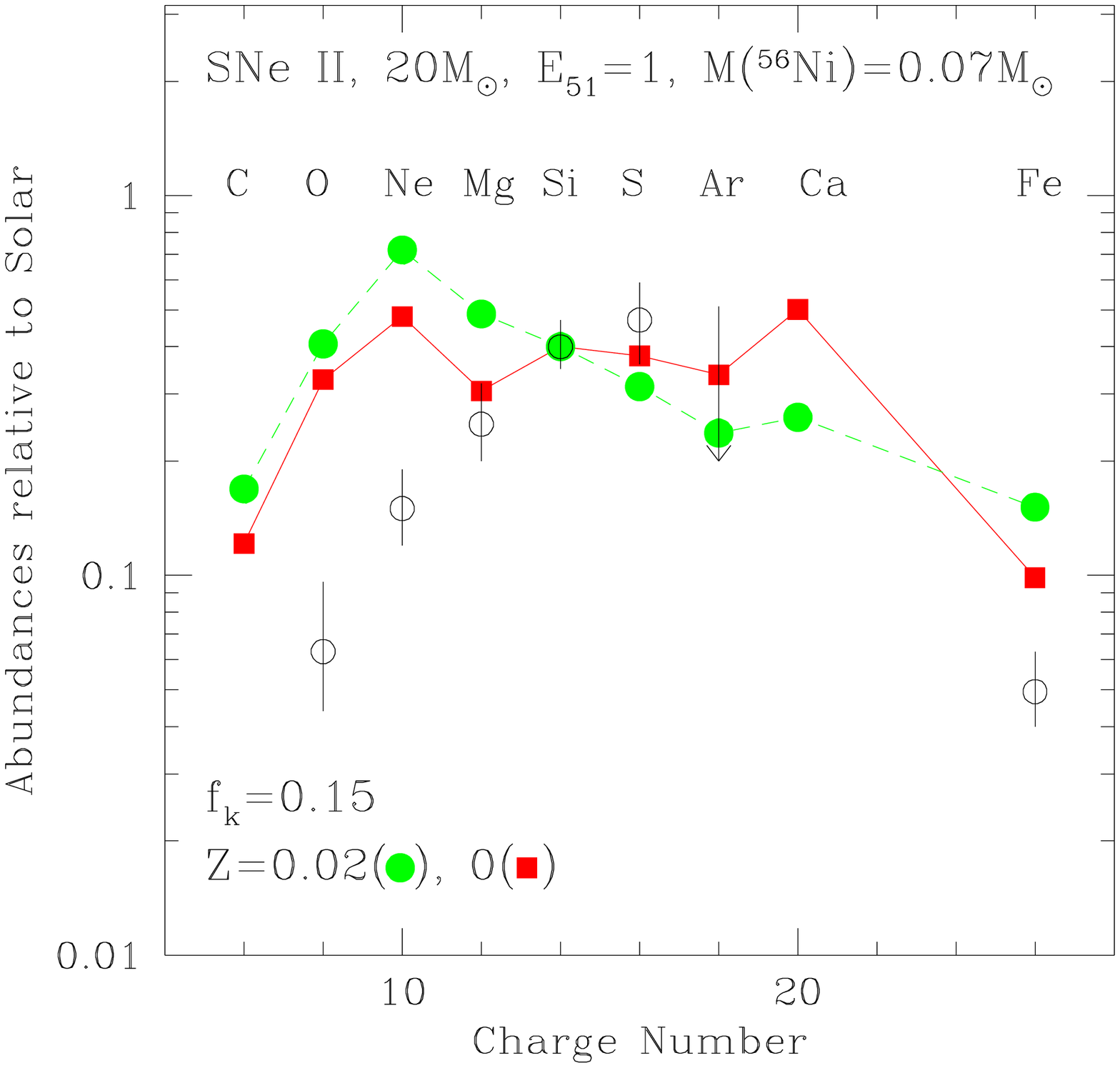}
\caption{Abundance patterns in the ejecta of 20$M_\odot$
SN II models for various metallicity compared with 
the observed abundances
(relative to the solar values) of M82. Here the left and
right panels show the cases with 
the relatively slow convective mixing ($f_k$=0.05) and
fast mixing ($f_k$=0.15), respectively.
The open circles with error bars show the M82 data. 
Theoretical abundances are
normalized to the observed Si abundance. \label{f1}}
\end{figure}

\placefigure{f1}

 From Figure 1, we find that these models are all have
much smaller Si/O ratios than the M82 data. 
The Si/O ratios are larger for $f_k=0.05$ than $f_k=0.15$ 
but still smaller than the observation
by a factor of $\sim$4. For other $E_{51}=1$ models in Table 1,
we find that [Si/O] $\lsim 0.57 $, being much 
smaller than the observation in all models. 
Generally, more massive core-collapse SNe ($M \lsim 130M_\odot$)
yield smaller Si/O ratios because the O yield
increases more rapidly with increasing mass than the Si yield.
Therefore, the Si/O ratio is likely to be largest for the smallest 
mass SN II. However, as shown in Table 1, the Si/O ratios for the
13$M_\odot$ SN II models are not large enough to be consistent 
with the M82 data. Contribution of even smaller mass SNe II 
would be negligible unless the Initial Mass Function was extremely steep.
Stars with $M \gsim 130M_\odot$
explode as PISNe, and the nucleosynthesis pattern may
be different. These cases are discussed in Section 4.
 This result is quite general, being consistent with 
our previous and other groups' results. We also note that
the Si/O ratio is
insensitive to the $^{12}$C($\alpha,\gamma)^{16}$O rate
(Nomoto et al. 1997a).

 It is interesting to note that, the 
abundance ratios for these even Z elements are insensitive to
the metallicity (Figure 1). This is because
the synthesis of these elements are mostly governed by the $\alpha$
- nuclei reactions, which are almost independent of metallicity
(Umeda et al. 2000). From this reason, in the rest of this paper,
we use the $Z$=0 models for the progenitor and supernova models.

\subsection{Hypernovae}

 Recent observations suggest that some
massive SNe  with $M \gsim 20-25M_\odot$ explode more than
ten times energetically than normal SNe II and SNe Ibc, which
may be called ``hypernovae''. The first discovery of such SNe
was SN1998bw (Galama et al. 1998; Iwamoto et al. 1998).
Then several other hypernovae and their candidates have been
discovered, namely, SNe 1997ef, 
1997cy, 1999as, and 2002ap (e.g., Nomoto et al. 2001; 
Mazzali et al. 2002).  

 Umeda et al. (2000) and Nakamura et al. (2001) have shown that the
large Si/O ratio may be the signature of the energetic supernova
explosions. Figure 2 compares the post-explosion abundance distribution
of the hypernova and the normal SN II models.
In the more energetic explosion,
the outer boundary of the explosive 
O burning region moves outwards. The explosive O burning 
to produce Si, S and Ar takes place 
when the peak temperature after the supernova shock passage 
exceeds $T_9 \simeq 4$, where $T_9$ is the peak temperature
in units of
10$^{9}$K. The region after the shock passage is radiation
dominant, so the peak temperature is approximately related to
the stellar radius $r$ and the explosion energy 
$E$ as $E \sim a T^4 4\pi r^3/3$, i.e.,  
$T_9 =E_{51}^{1/4} (r/3.16\times 10^{4}$ km)$^{-3/4}$.
For larger $E$, therefore, the O burning region extends 
to larger $r$.
On the other hand, the outer edge of the O-rich region is 
fixed at the C+O/He interface, where the density is too low
for explosive burning. This is why the Si/O ratio
is enhanced for more energetic explosions.

\begin{figure}
\plotone{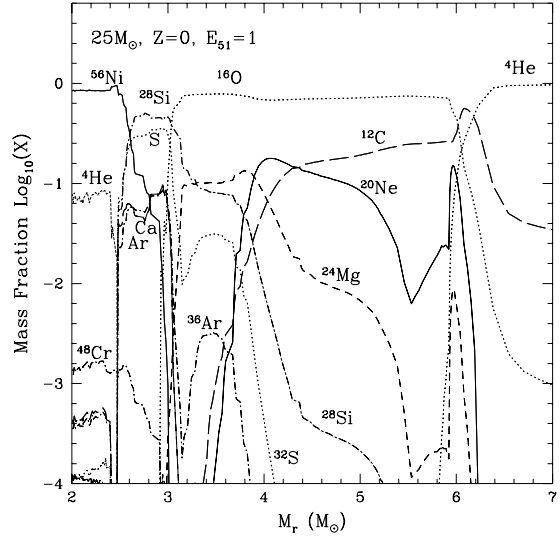}
\plotone{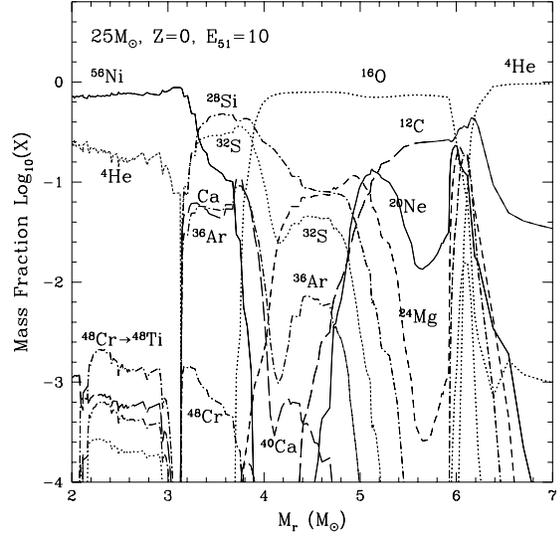}
\caption{Abundance distribution after the supernova explosion
of normal ($E_{51}$=1) and energetic ($E_{51}$=10) core-collapse 
SNe ($f_k=0.05$ models).
The unstable isotope $^{48}$Cr decays into $^{48}$Ti.
\label{f2}}
\end{figure}

\placefigure{f2}

 Figure 3 shows
the nucleosynthesis pattern of the 25$M_\odot$ ($Z$=0, $f_k=0.05$) 
models for various
explosion energies, $E_{51}$=1, 10 and 30. Here the abundances are
normalized to the observed Si abundance and the mass-cuts 
(i.e., the compact remnants' masses) are chosen to 
eject the Fe mass to fit the data point. Thus the ejected Fe masses
are 0.07, 0.095 and 0.12 $M_\odot$, for $E_{51}$=1, 10 and
30, respectively. 
The mass-cuts for these models are 2.3, 3.1 and 4.0 $M_\odot$,
respectively. Table 2 summarize
the ejected masses of O, Ne, Mg, Si, S and Fe as a function of
the explosion energy for the 25 and 30 $M_\odot$ ($Z$=0) models.

\begin{figure}
\plotone{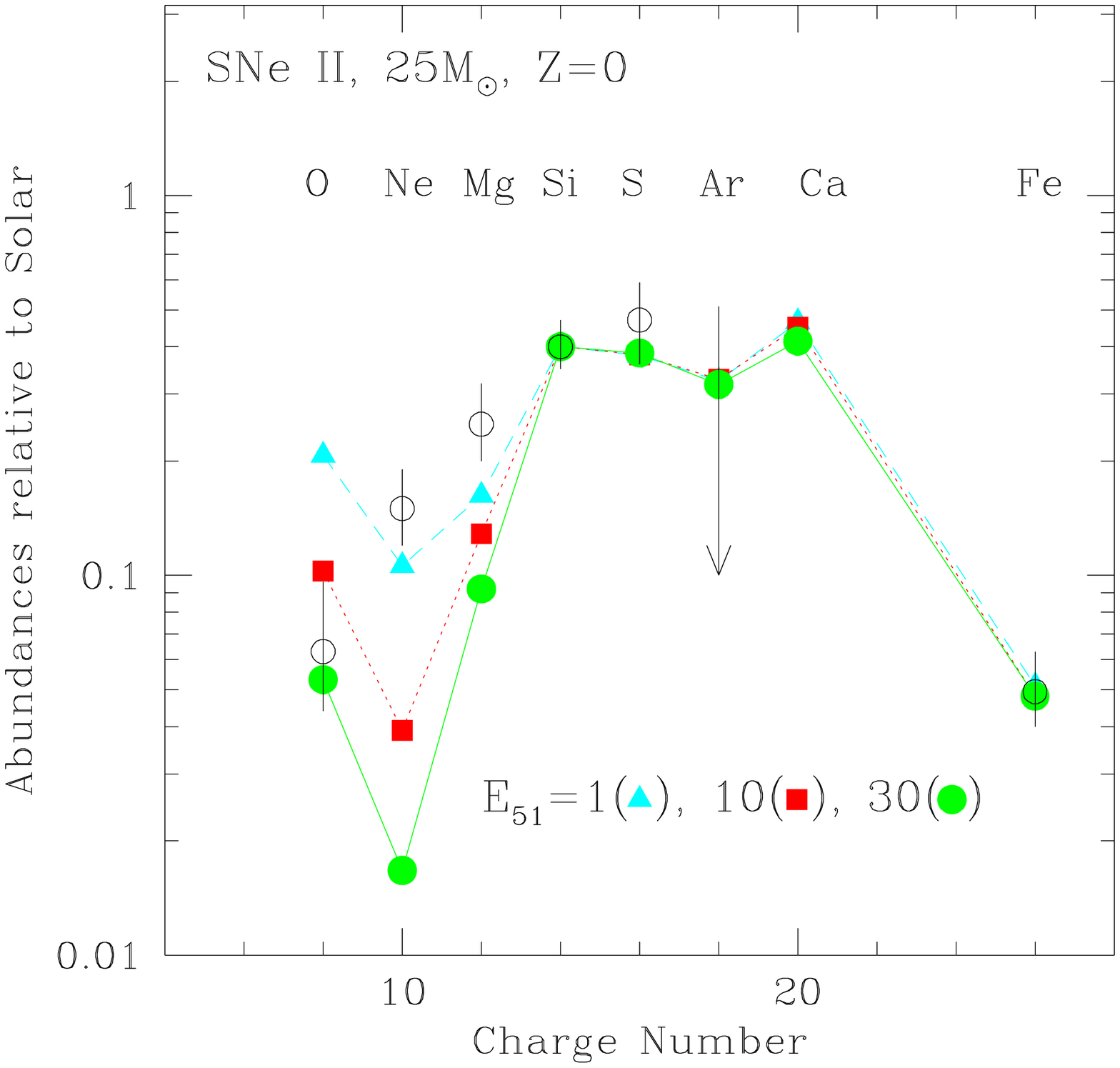}
\plotone{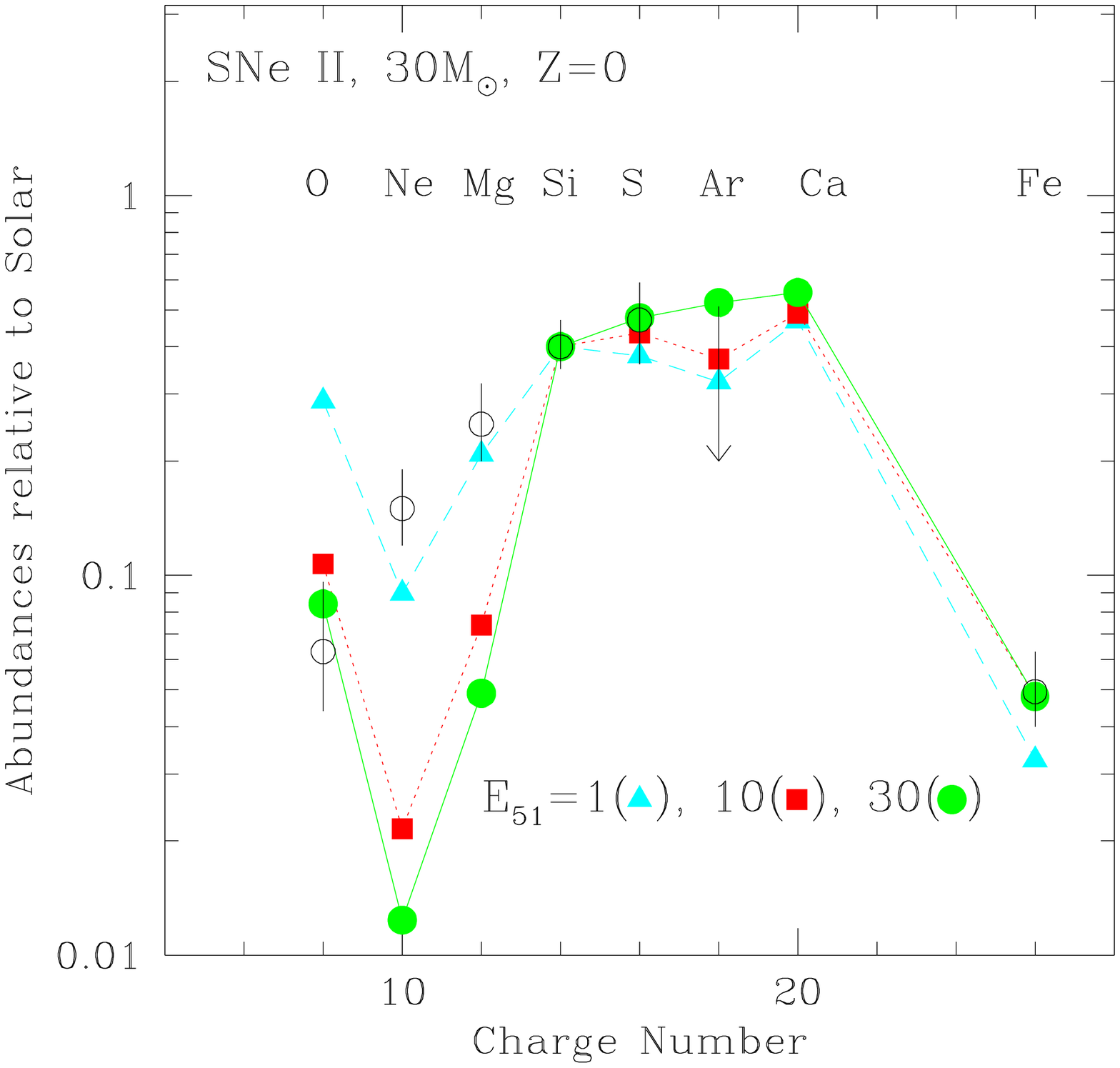}
\caption{Same as Figure 1 but for 25 and 30$M_\odot$
Pop III SN II with normal explosion energy $E_{51}=1$ and 
hypernova models with $E_{51}=$ 10, 30 ($25M_\odot$) 
and  $E_{51}$= 30, 50 ($30M_\odot$). \label{f3}}
\end{figure}

\placefigure{f3}

 The observational and theoretical mass ratios of Si/O
are compared in the top panel of Figure 4.
The observed large abundance ratio of Si/O
can be reproduced for $E_{51} \gsim$ 10 in the 25$M_\odot$ models,
and $E_{51} \gsim$ 30 in the 30$M_\odot$ models.
Here, the results are for the $f_k=0.05$ models. For the
$f_k=0.15$ models, which is more O-rich, higher energy 
is required to be consistent with observations.

 Similar Si/O ratios can be obtained for more massive stars if the
explosion energy is sufficiently large. Therefore, more massive
core-collapse hypernovae with initial masses 
of $\sim 50 - 100M_\odot$ could
also be consistent with the M82 abundance pattern, although 
we do not know how much fraction of these massive stars actually
explode rather than collapsing to black holes without explosion.

\begin{figure}
\epsscale{1.3}
\plotone{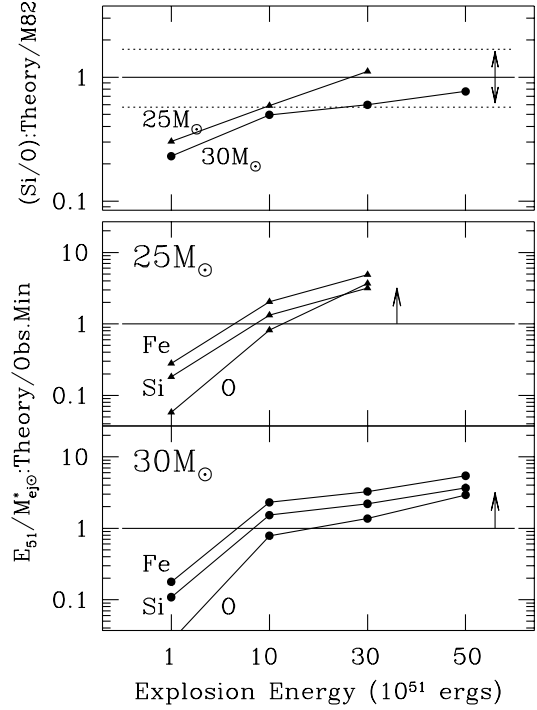}
\caption{ Comparison of the theoretical Si to O 
mass fraction ratios with the observed range, which is shown by 
the arrows, as a function of explosion
energy (top panel). Comparison of the theoretical energy to 
heavy element (Fe, Si and O) mass ratios
with the observed lower limits as a function of the explosion
energy (middle panel for the 
25$M_\odot$ models and bottom panel for the 30$M_\odot$
models, respectively). 
Here, shown are the $f_k=0.05$ models. For the
$f_k=0.15$ models, which are more O-rich, higher energy 
is required to be consistent with observations.
\label{f4}
}
\end{figure}

\placefigure{f4}

 In these results, Ne and Mg appear to be underproduced.
However, the ratios (Ne, Mg)/O are sensitive to the
uncertain reaction rate $^{12}$C($\alpha,\gamma)^{16}$O 
and convective parameter $f_k$ as seen in Figure 1.
Thus a better agreement with observations can be obtained by
choosing appropriate parameters, which does not change the
present conclusion on the Si/O ratio.
 
\section{Energy to Heavy Element Mass Ratio}

 In this section, we discuss another observational constraint which
theoretical models should satisfy.
 Suppose that the hot plasma in M82 is a
mixture of the SN ejecta and the interstellar gas, and
most of the thermal energy of the hot plasma
was supplied by supernovae. Then
we can constrain the ratio of the energy released by one 
supernova, $E$,  to the mass of the heavy elements 
as follows (Tsuru et al. 1997).  

 The thermal energy of the hot plasma supplied by one SN explosion can be
written as
\begin{equation}
\frac{3}{2}\frac{M_{\rm ej}+M_{\rm am}}{\mu m_{\rm H}}
k_{\rm B} T \lsim E,
\end{equation}
where $M_{\rm ej}$ and $M_{\rm am}$ are the masses of the
ejecta and the ambient gas heated by a single SN and,
$\mu (\simeq 0.6)$ is the mean molecular weight. The ratio of the
mass of each heavy element ($M^*$) to the total mass ($M_{\rm total}$) is
expressed as
\begin{equation}
\frac{M^*}{M_{\rm total}}= 
\frac{M^*_{\rm ej}+M^*_{\rm am}}{M_{\rm ej}+M_{\rm am}}
\gsim \frac{M^*_{\rm ej}}{M_{\rm ej}+M_{\rm am}},
\end{equation}
where the quantities with asterisk denote the mass of each heavy element.
 From these equations, we obtain: 
\begin{equation}
\frac{E_{51}}{M^*_{\rm ej \odot}} \gsim 4.9\times 10^{-3}
\left( \frac{k_{\rm B} T}{1 {\rm keV}} \right) 
\left( \frac{M^*}{M_{\rm total}}\right )^{-1}
\equiv \left( \frac{E_{51}}{M^*_{\rm ej \odot}} \right)_{\rm min},
\end{equation}
where $M^*_{\rm ej \odot}$ is the ejected mass 
in units of $M_\odot$ of each heavy element synthesized by the SN. 
 For a given heavy element mass in the ejecta,
this inequality determines the minimum explosion energy required for
the supernova model to be consistent with the observations. 

Using this relation and the observed values of k$_{\rm B} T (\simeq 1$ keV)
and $M^*/ M_{\rm total}$ (Tsuru et al. 1997 and references therein),
the minimum energy to the heavy element mass ratios for O, Si, and Fe
are given as follows: 

\begin{equation}
(E_{51}/M^*_{\rm ej \odot})_{\rm min}(\rm{O})
=7.9^{+3.5}_{-2.7}, 
\end{equation}

\begin{equation}
(E_{51}/M^*_{\rm ej \odot})_{\rm min}(\rm{Si})
=17.5^{+2.4}_{-2.7}, 
\end{equation}

\begin{equation}
(E_{51}/M^*_{\rm ej \odot})_{\rm min}(\rm{Fe})
=51.2^{+11.4}_{-11.5}.
\end{equation}
The limits for other elements can also be
obtained, but these three elements are sufficient for constraining
$E_{51}$.

Here, we have neglected the contribution of the stellar wind 
to the total energy of the hot plasmas. The energy of stellar
wind from solar metallicity stars is estimated to be $\sim 0.2 \times
10^{51}$ erg from the observation of OB associations (Abbott 1982).
This is much smaller than the typical explosion energy of a normal SN II,
$\sim 10^{51}$ erg. This energy is even smaller for a lower metal environment
as in M82, because the wind mass-loss rate decreases with metallicity
as $\propto Z^{0.8}$ (Leitherer, Robert \& Drissen 1992).
Furthermore, we are considering the energy from a hypernova with 
$\gsim 10^{52}$ erg. Therefore, it is not a bad approximation 
to neglect the energy from stellar winds in the hot-plasma. 

 In the middle and bottom panels of Figure 4, we compare the observed
ratios ($E_{51}/M^*_{\rm ej \odot})_{\rm min}$ with the theoretical
$E_{51}/M^*_{\rm ej \odot}$.
For the $E_{51}=1$ models, the ratios $E_{51}/M^*_{\rm ej \odot}$
for all elements are much smaller than the minimum values in Eq. (3).
On the other hand, all other 
energetic models shown here satisfy the constraints.
Therefore, the 25$M_\odot$ models with $E_{51}\gsim 10$ and
the 30$M_\odot$ models with $E_{51}\gsim 30$ are consistent with
both constraints on
the Si/O abundance ratio and the energy to mass ratio.
Here, we note again that these results are for the $f_k=0.05$ 
models and larger energy is required for $f_k=0.15$ models
to be consistent with observations.

\section{Pair Instability Supernovae}

 It has been known that the stars with $M \sim 150-300 M_\odot$
also explode energetically as pair instability SNe (PISNe) 
(e.g., Barkat, Rakavy \& Sack 1967; Fraley 1968; 
Ober, El Eid \& Fricke 1983; Woosley \& Weaver 1982). 
During oxygen burning the cores of
these stars enter into the electron - positron pair creation region.
Then the adiabatic index of these cores becomes less than 4/3
and the stars collapse rapidly. The central 
temperature increases during the collapse and rapid O-burning
takes place to produce a large amount of thermal energy and Si. 
For the stars with $M \lsim 300M_\odot$ the thermal energy produced
by oxygen burning exceeds the gravitational potential energy,
and the stars disrupt completely. Compared with normal SNe II, 
the explosion energies are
larger, typically $2-4  \times ~10^{52}$ erg, and the Si abundance
in the ejecta is also larger (see Table 2). 
During the PISN explosion a
large amount of Fe ($\sim 10 M_\odot$) can be produced (Umeda \& Nomoto
2002; Heger \& Woosley 2002), although
the amount of Fe production is very 
sensitive to the maximum temperature attained.

In Figure 5, we show the abundance distribution of the 200$M_\odot$
PISN model. 
In Table 2, the ejected masses of O, Ne, Mg, Si, S and Fe are shown 
for the 170 and 200$M_\odot$ Pop III PISN (Umeda \& Nomoto 2002).
The abundance patterns of the 170 and 200$M_\odot$ stars are 
compared with the M82 data in Figure 6.
It seems that the large [Si/O] ratio in M82
can be realized by PISN nucleosynthesis.

\begin{figure}
\epsscale{1}
\plotone{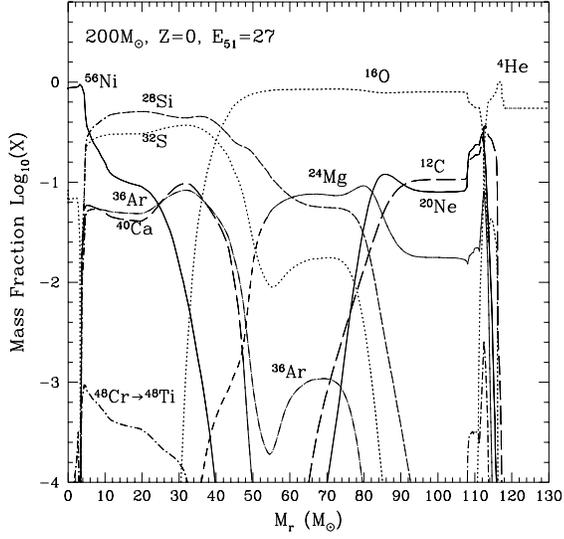}
\caption{Same as Figure 2 but for 200$M_\odot$
Pop III PISN model. \label{f5}}
\end{figure}

\placefigure{f5}

\begin{figure}
\plotone{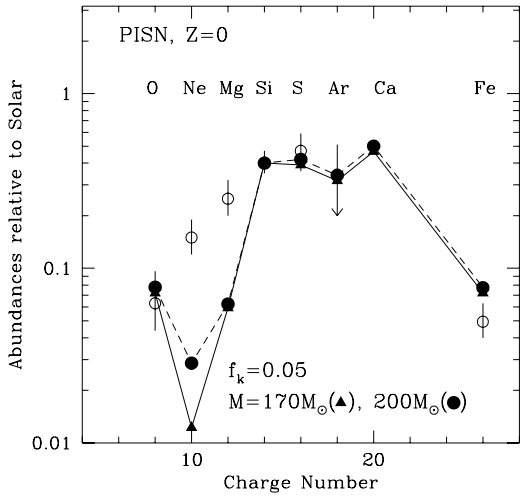}
\caption{Same as Figure 1 but for 170 and 200$M_\odot$
Pop III PISN models. \label{f6}}
\end{figure}

\placefigure{f6}

 However, we also need to compare the energy to mass ratio for
the PISN models.
 Since the amount of ejected O is least model 
dependent, we show the energy to the O mass ratio for the PISNe.
Our models with  $M$= 170 and 200$M_\odot$ ($Z$=0) have 
 $E_{51}/M^*_{\rm ej \odot}$(O)= 22.4/44.2 = 0.51 
and 26.8/56.0 = 0.48, respectively. These are much smaller than the
observed lower limit of 7.9 (Eq. 4), because the ejected O mass is
more than ten times larger than in the core-collapse SNe.

 This ratio may be larger for more extremely massive stars.
For example, in the model of Heger \& Woosley (2002),
their 130$M_\odot$ He star model, corresponding to the initial stellar mass
of 260$M_\odot$, gives
$E_{51}/M^*_{\rm ej \odot}$(O)= 87.1/29.9 = 2.9.
For these models, however, the O/Fe ratios are typically too small 
([O/Fe]=$-1.1$ for the  130$M_\odot$ He star model) 
to be compatible with observations.
 
\section{Conclusions and Discussion }

 X-ray observations have shown that the chemical abundance in the
starburst galaxy M82 is quite rich in Si and S compared with
oxygen. Such an abundance pattern cannot be explained with any
combination of conventional Type I and II supernova yields. Also the
energy to heavy element mass ratio of the observed hot plasma 
is much higher than that of normal supernovae.
We have calculated explosive nucleosynthesis in core-collapse hypernovae
to show that the abundance pattern and the large energy of the hot
plasma of M82 can be explained with such hypernovae as
($M$, $E_{51}$) = (25, 10) and (30, 30).  More massive and more
energetic core-collapse explosions 
would also satisfy the observed constraints.
We have also investigated pair-instability supernovae ($M \sim 150-300
M_\odot$) and conclude that the energy to heavy element mass ratio in
these supernovae is too small to explain the observation.

 Such ``hypernova'' (energetic core-collapse SN) 
explosions are expected to occur for stars more massive
than $\gsim 20-25 M_\odot$ (Mazzali et al. 2002). 
The upper mass limit of core-collapse
SNe is still uncertain. Too massive stars may collapse to 
black holes without explosion. 

The question is how the abundance in M82 can be dominated by such
hypernovae. One possible explanation
is the age effect. As mentioned in Introduction, 
the age after the beginning of
star-burst is estimated to be $\sim 10^6 - 10^7$ years. 
On the other hand, the
lifetime of our Pop III stars are 1.40, 1.10, 0.81
and 0.70 $\times 10^{7}$ years for 15, 20, 25 and 30$M_\odot$ models,
respectively (Umeda et al. 2000).  This is consistent with the
assumption that only such massive stars as
$M \gsim 20-25M_\odot$ have exploded in M82.

 There have been some suggestions that hypernovae might be correlated with
gamma-ray bursts (GRBs) (e.g., Iwamoto et al. 1998). 
However, how much fraction
of hypernovae is associated with GRBs is still unknown. Paczynski (2001)
discussed that the SN rate (including the energetic SNe) is much higher than 
the GRB rate. On the other hand, we have shown that in order to 
explain the observed abundance pattern
of M82, dominant fractions of
massive stars ($M > \sim 25 M_\odot$) needs to be hypernovae.
This implies that hypernovae are much more frequent than GRBs.
If GRBs are associated with hypernovae,
it is likely that they occur only for certain special cases of hypernova
explosions. For example, the explosion energy needs to exceed
a certain value, the explosion needs to be extremely aspherical, i.e.,
the ejecta needs to be strongly beamed, and
both hydrogen and helium envelopes need to be stripped off before
the explosion.
In fact, GRB980425 associate with SN1998bw is the exceptionally
weak GRB (Galama et al. 1998), while SN1998bw is among the most
energetic hypernovae so far (Nomoto et al. 2001; Mazzali et al. 
2002).

 What is the implication of this work to the intermediate mass
black holes recently discovered in M82?  One possible scenario
for the formation of such black holes is the merging of massive stars
in the dense stellar clusters (Portegies et al. 1999).
This scenario is consistent with the results of this paper, because
such merging increases the number of massive star explosions. 
However, our results constrain the merging history. The typical mass
range for the merged stars cannot be in the range for
PISN ($M \sim 150-300M_\odot$) to be consistent with the energy to heavy
element mass ratios observed in M82.

\bigskip

This work has been
supported in part by the grant-in-Aid for Scientific Research
(12640233, 14047206, 14540223) of the Ministry of 
Education, Science, Culture, Sports, and Technology in Japan.


\clearpage

\clearpage

\begin{deluxetable}{lccccccccccc}
\tabletypesize{\scriptsize}
\tablecaption{Ejected masses of selected elements 
in $M_\odot$ for various metallicity $Z$ and mass $M$
in SN II models with explosion energies $E_{51}=1$.
\label{tbl-1}}
\tablewidth{0pt}
\tablehead{ & \multicolumn{7}{c}{Ejected Mass ($M_\odot$) 
 }}
\startdata 
($M$, $Z$) & C& N& O & Ne & Mg & Si & S & Ar& Ca& Fe  & [Si/O] \cr 
(13, 0.02)&0.11&0.058&0.26&0.037&0.025&0.072&0.039&0.0074&0.0059&0.091& 0.57\cr 
(13, 0) &0.19&0.10&0.22&0.026& 0.023 & 0.049 & 0.026&0.0051&0.0048&0.070& 0.56\cr 
(15, 0.02)&0.16&0.060&0.39&0.058 &0.035&0.085&0.044&0.0081&0.0066&0.094& 0.47 \cr 
(15, 0)&0.19&0.014&0.53& 0.072& 0.030& 0.11&0.063&0.013&0.013& 0.070  & 0.43 \cr 
 \hline 
(20, 0.01)&0.31&0.028&1.76&0.12&0.076&0.28&0.15&0.026&0.022&0.082& 0.33  \cr 
(20, 0.004)&0.33&0.015&1.27&0.10&0.064&0.23&0.13&0.024&0.021&0.081& 0.39  \cr 
(20, 0.001)&0.29&0.0034&1.65&0.14&0.077&0.28&0.17&0.032&0.030&0.073& 0.36  \cr 
(20, $10^{-4}$)&0.26&4.8e-4&1.27&0.18&0.080&0.28&0.16&0.028&0.024&0.071&0.48\cr 
(20, $10^{-5}$)&0.26&0.0018&2.02&0.15&0.098&0.33&0.21&0.041&0.040&0.070& 0.35\cr 
(20, 0)&0.26&2.7e-4&1.56&0.12&0.071&0.25&0.15&0.030&0.030&0.070 & 0.34  \cr 
 \hline 
(25, 0.02)&1.01&0.050&5.79&0.66&0.22&0.67&0.30&0.049&0.032&0.11 & 0.20  \cr 
(25, 0)&0.61&3.6e-4&2.18&0.20&0.12&0.31&0.17&0.033&0.031&0.070& 0.28  \cr 
(30, 0.02)&0.96&0.061&3.43&2.31&0.68&0.25&0.091&0.017&0.0083&0.12&$-0.01$  \cr 
(30, 0)&0.36&1.9e-4&4.80&0.27&0.24&0.49&0.27&0.052&0.050& 0.070 & 0.14  \cr 
(40, 0.02)&0.81&0.086&6.41&2.06&0.48&0.36&0.16&0.026&0.019&0.12 & $-$0.12  \cr 
 \hline 
(20, 0.02)\tablenotemark{a}& 0.27 &0.048
& 2.03 & 0.65 & 0.17 & 0.15 & 0.068& 0.011& 0.0084&0.10 & $-$0.01  \cr 
(20, 0)\tablenotemark{a}& 0.21 &9.9e-4
& 1.77 & 0.47 & 0.11 & 0.16 & 0.089& 0.018& 0.017&0.071 & 0.09  \cr 
(50, $10^{-4}$)\tablenotemark{b}&5.89&0.0015 
& 11.6 & 0.38 & 0.21 & 0.37 & 0.17&0.030&0.027& 0.070 & $-$0.36   \cr 
(50, $10^{-4}$)\tablenotemark{c}& 1.14 & 9.7e-4 
& 12.5& 2.41 & 0.79 & 0.95 & 0.45 & 0.077&0.068& 0.071 & 0.01   \cr 

\enddata
\tablecomments{The mass of Fe ($^{54}$Fe, $^{56}$Fe, $^{57}$Fe)
depends on the mass-cut. Here, the mass-cut is
chosen to eject 0.07 $M_\odot$ of $^{56}$Ni, which decays
into $^{56}$Fe, via $^{56}$Co. Except for the
last four items, the adopted
parameters are $f_k$=0.05 and 1.4 $\times$ Caughlan \& Fowler 
(1988; CF88 hereafter) 
for the $^{12}$C($\alpha,\gamma)^{16}$O rate.}
\tablenotetext{a}{$f_k$=0.15, 1.3 $\times$ CF88 for the 
$^{12}$C($\alpha,\gamma)^{16}$O rate}
\tablenotetext{b}{$f_k$=0.1, CF88 $^{12}$C($\alpha,\gamma)^{16}$O rate}
\tablenotetext{c}{$f_k$=0.5, CF88 $^{12}$C($\alpha,\gamma)^{16}$O rate}
\end{deluxetable}


\begin{deluxetable}{lccccccccccc}
\tabletypesize{\scriptsize}
\tablecaption{Same as Table 1 but for the $Z$=0, 
$M$=25 \& 30 $M_\odot$ core-collapse SNe
with various explosion energies. 
\label{tbl-2}}
\tablewidth{0pt}
\tablehead{ & \multicolumn{7}{c}{Ejected Mass ($M_\odot$) 
 }}
\startdata 
($M$, $E_{51}$) & C& N& O & Ne & Mg & Si & S &Ar&Ca& Fe & [Si/O] \cr 
(25, 1) &0.61&3.6e-4&2.18&0.20&0.12&0.31&0.17&0.033&0.031&0.07& 0.28 \cr 
(25, 10)&0.40&7.2e-4&1.55 & 0.11&0.13&0.43&0.24&0.048&0.044&0.095 & 0.58 \cr 
(25, 30)&0.17&0.0023&1.03&0.059&0.12&0.54&0.30&0.060&0.052&0.12& 0.85 \cr 
 \hline 
(30, 1)&0.36&1.9e-4&4.80&0.27&0.24&0.52&0.27&0.052&0.050&0.11&0.17 \cr 
(30, 20)&0.24&2.2e-4&3.21 &0.14 &0.15 & 0.75&0.44&0.079&0.074&0.16&0.50\cr 
(30, 30)&0.17&1.4e-4&2.77&0.10&0.13&0.78&0.51&0.093&0.081& 0.16 & 0.58  \cr 
(30, 50)&0.11&9.1e-5& 2.16&0.058&0.086&0.78&0.56&0.13&0.092&0.18  & 0.69 \cr 
 \hline 
(170, 22.4)&2.30&0.010&44.2&1.19&1.94&16.2&8.06&1.42&1.32&3.6 & 0.69  \cr 
(200, 26.8)&4.24&5.8e-4&56.0&3.75&3.08&21.2&13.1&2.36&2.32&7.2 & 0.71  \cr 
\enddata
\tablecomments{Here, the mass-cut (or Fe mass)
is chosen to satisfy the observed Si/Fe ratio.
The $M$=170 \& 200 $M_\odot$ PISN models are also shown.
The SN II models with $E_{51}=1$ and PISN models do not
satisfy the observational constraints Eq.s (4-6), but 
the core-collapse hypernova models ($E_{51}\geq 10$) shown 
here satisfy the constraints. The adopted
parameters are $f_k$=0.05 and 1.4 $\times$ CF88
for the $^{12}$C($\alpha,\gamma)^{16}$O rate.}
\end{deluxetable}

\clearpage

\end{document}